\documentclass[]{jfm}
\usepackage{graphicx}
\usepackage{newtxtext}
\usepackage{newtxmath}
\usepackage{natbib}
\usepackage{hyperref}
\usepackage{dcolumn}
\usepackage{bm}
\usepackage{gensymb}
\usepackage{tabularx}
\usepackage{color}
\usepackage{braket}
\usepackage{comment}
\hypersetup{colorlinks=true,urlcolor=blue,citecolor=blue}

\title{Mode interaction in inclined-layer convection:\\ a physical interpretation}
\shorttitle{Mode interaction in inclined-layer convection: a physical interpretation}
\shortauthor{Z. Zheng, L.S. Tuckerman and T.M. Schneider}
\author{Zheng Zheng\aff{1,2}, 
Laurette S. Tuckerman\aff{3}\corresp{\email{laurette.tuckerman@espci.fr}} \and 
Tobias M. Schneider\aff{2}}

\affiliation{
\aff{1} School of Mechanical Engineering, Southeast University, 211189 Nanjing, China
\aff{2} Emergent Complexity in Physical Systems Laboratory (ECPS), \'Ecole Polytechnique F\'ed\'erale de Lausanne, CH 1015 Lausanne, Switzerland
\aff{3} Physique et M\'ecanique des Milieux H\'et\'erog\`enes (PMMH), CNRS, ESPCI Paris, PSL University, Sorbonne Universit\'e, Universit\'e de Paris, 75005 Paris, France}

\begin{document}
\maketitle

\begin{abstract}
	We carry out a numerical investigation of the Oberbeck--Boussinesq equations in a fluid layer inclined against gravity. Equilibria containing different numbers of co-rotating rolls are computed. Mixed-mode branches connect branches containing states with four and six identical co-rotating rolls. These branches display two contrasting routes to reducing (or increasing) the number of rolls, one in which rolls disappear (or appear) and another route in which rolls merge (or split). These results are interpreted in light of the $(2,3)$ mode interaction scenario. 
\end{abstract}


\section{Introduction}
The study of mode interactions in pattern-forming systems is a classic field of fluid mechanics \citep{crawford1991symmetry}. Even the one-dimensional problem -- that of wavelength selection -- can be complicated and can display interesting phenomena. The prototype is a homogeneous system which undergoes a sequence of supercritical bifurcations to modes which are necessarily trigonometric. These bifurcations produce primary branches, which then undergo secondary bifurcations to mixed-mode branches. This is the essence of the Eckhaus instability \citep{eckhaus1965studies, tuckerman1990bifurcation} and of steady-state mode interaction.

The theoretical basis for these scenarios in terms of model equations, more specifically normal forms, is well known \citep{dangelmayr_Armbruster_1986, dangelmayr1986steady, golubitsky1985singularities, golubitsky1988singularities}. In addition, many experiments and fluid-dynamical simulations have displayed manifestations of these phenomena in a variety of configurations: convection rolls, but also Taylor-Couette flow \citep{dominguez1986eckhaus}, the shear-driven cavity \citep{bengana2019bifurcation}, and many others. However, there have been few attempts to relate model equations to full hydrodynamic simulations. Here, we carry out an investigation of inclined-layer convection (ILC) in which states with four and six co-rotating rolls compete. We show that the results of these simulations match those of the normal form equations concerning $(2,3)$ interaction. 

In ILC, a wall-bounded fluid layer is inclined against gravity and is heated from one wall and cooled from the other \citep{Daniels2000, Subramanian2016}. Unlike pure shear flows whose base state remains theoretically linearly stable for all Reynolds numbers, thermal convection admits linear instability. In our case, the primary instability is shear driven and the resulting secondary state consists of transverse rolls. The rest of the paper is structured as follows. In \S \ref{sec:methods_DNS}, we briefly present the numerical methods used in our research. In \S \ref{sec:connector}, we study a connector branch exhibiting the roll-merging and roll-disappearing phenomenon, and analyse the $D_2$--$D_3$ mode interaction problem. Concluding remarks will be outlined in \S \ref{sec:conclusion}.

\section{Inclined-layer convection system and numerical methods}
\label{sec:methods_DNS}
We refer readers to \citet{Reetz2020a, Reetz2020b} as well as \citet{Zheng2024part1, Zheng2024part2, Zheng2025PhD} for more detailed descriptions of the numerical methods used in this research. Here, we will only summarize the key points. The computations are performed with the ILC extension module of the pseudo-spectral code \textit{Channelflow 2.0} \citep{Gibson2019}. Our computational domain is of size $[L_x, L_y, L_z] = [12, 6, 1]$, and is discretized by $[N_x, N_y, N_z] = [96, 48, 31]$ collocation points. Although our computations were performed in three dimensions as part of a larger investigation, the phenomena described in this paper are independent of the $y$ direction and have no velocity in the $y$ direction, so we restrict further descriptions to the $(x,z)$ plane.

\begin{figure}
	\centering
	\captionsetup{font={footnotesize}}
	\captionsetup{width=13.5cm}
	\captionsetup{format=plain, justification=justified}
	\includegraphics[width=\columnwidth]{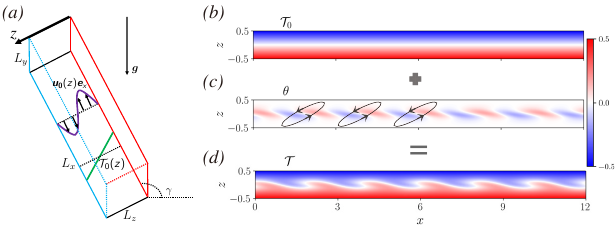}
	\caption{\label{ILC_figure}(a) Computational domain. A differentially heated fluid layer is inclined against gravity at an angle of $\gamma=100\degree$. The cavity is bounded between walls at $z=\pm0.5$, oriented at angle $\gamma$ from the horizontal, and the domain is periodic in the $y$ and $z$ directions. Velocity and temperature profiles of the conductive base solution are shown as the purple curve and the green line. (b) Temperature $\mathcal{T}_0$ of the conductive state, (c) temperature deviation $\theta \equiv \mathcal{T} - \mathcal{T}_0$, and (d) total temperature field $\mathcal{T}$ of the six-roll state at $Ra=12000$. Convection rolls are schematically delineated by ovals, with arrows giving the direction of rotation and colours indicating temperature.}
\end{figure}

The Oberbeck--Boussinesq equations governing our system are 
\begin{subequations}
	\begin{align}
		\dfrac{\partial \boldsymbol u}{\partial t}+ (\boldsymbol u \cdot \nabla)\boldsymbol u &= -\nabla p + \left(\frac{Pr}{Ra}\right)^{1/2} \nabla^2 \boldsymbol u - \hat{\boldsymbol{g}}\mathcal{T}, \label{appa} \\
		\dfrac{\partial \mathcal{T}}{\partial t} + (\boldsymbol u \cdot \nabla)\mathcal{T} &= \left(\frac{1}{Pr\: Ra}\right)^{1/2} \nabla^2 \mathcal{T}, \label{appb} \\
		\nabla \cdot \boldsymbol u&= 0, \label{appc} 
	\end{align}
\end{subequations}
where $\boldsymbol u=(U,W)$, $p$, $\mathcal{T}$ represent the velocity, pressure and temperature fields respectively. The computational domain is bounded in the wall-normal direction at $z=\pm 0.5$ where $\boldsymbol u(z=\pm0.5)=0$ and $\mathcal{T}(z=\pm0.5)=\mp0.5$, and is inclined against gravity by $\gamma$: $\hat{\boldsymbol{g}}= -\sin(\gamma)\boldsymbol e_x - \cos(\gamma)\boldsymbol e_z$. The basic conductive solution is: 
\begin{subequations}
	\begin{align}
		\boldsymbol u_0(z) &= \dfrac {\sin(\gamma)\sqrt{Ra}}{6\sqrt{Pr}} \left(z^3 - \frac{1}{4}z \right) \boldsymbol e_x, \\
		\mathcal{T}_0(z) &= -z, \\
		p_0(z) &= \Pi - \cos(\gamma)z^2/2,
	\end{align}
\end{subequations}
with arbitrary pressure constant $\Pi$. Figure \ref{ILC_figure}(a) shows the computational domain and the temperature and velocity profiles of the conductive state.
The ILC system has the following symmetries:
\begin{subequations}
	\begin{eqnarray}
		&\pi_{xz}[U,W,\mathcal{T}](x,z) \equiv [-U,-W,-\mathcal{T}](-x,-z), \\
		&\tau(\Delta x)[U,W,\mathcal{T}](x,z) \equiv [U,W,\mathcal{T}](x+\Delta x,z).
	\end{eqnarray}
\end{subequations}
These operations generate the group $\braket{\pi_{xz}, \tau(\Delta x)} \sim [O(2)]_{xz}$. In this work, we set the Prandtl number to $Pr=1.07$, the inclination angle to $\gamma=100\degree$, and we investigate Rayleigh numbers in the range $9000 \leq Ra \leq 12000$.

We compute equilibria using the shooting-based Newton method implemented within \textit{Channelflow 2.0} \citep{Gibson2019}. Flow fields are visualized via the deviation $\theta \equiv \mathcal{T} - \mathcal{T}_0$ of the temperature from the conductive solution and its $L_2$-norm $\|\theta\|_2$ is used as an observable for plotting bifurcation diagrams. Figure \ref{ILC_figure}(b) shows the temperature of the conductive state, which is constant in $x$ and linear in $z$. Below it, figure \ref{ILC_figure}(c) shows the deviation field computed for $Ra=12000$ and figure \ref{ILC_figure}(d) shows the total temperature field which is the sum of the two. With an inclination angle of $\gamma=100 \degree$, the configuration is close to that of vertical convection: the instability is driven by shear (here $du_0/dz$) in contrast to the buoyancy-driven counter-rotating rolls that occur in Rayleigh--B\'enard convection. The rolls in the deviation field each contain one red region, indicating hot rising fluid, and one blue region, indicating cold falling fluid. These are {\it co-rotating} rolls, because all of them rotate in the same direction and they are also called {\it transverse} rolls because the $y$ orientation of their axes is transverse to the $x$ direction of the main flow. The flow in figure \ref{ILC_figure}(c) contains six rolls; the changes in the number of these rolls is the topic of our investigation.

\section{Rolls and connectors}
\label{sec:connector}
\subsection{Overview}
\begin{figure}
	\centering
	\captionsetup{font={footnotesize}}
	\captionsetup{width=13.5cm}
	\captionsetup{format=plain, justification=justified}
	\includegraphics[width=\columnwidth]{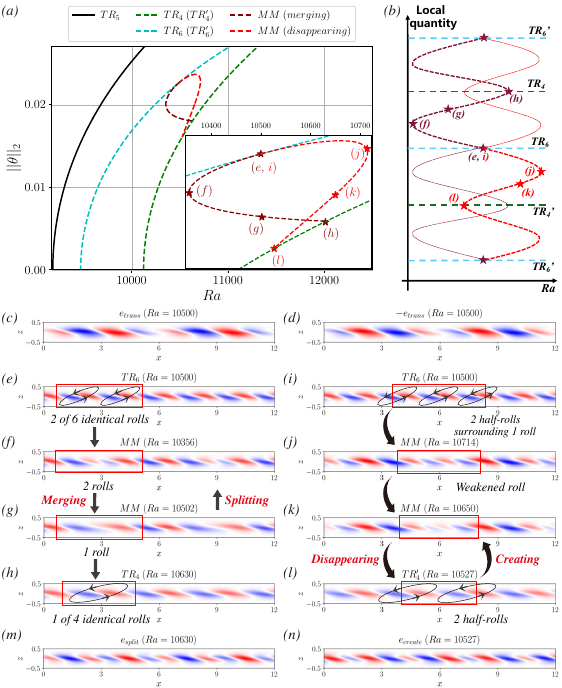}
	\caption{\label{BD_connector}(a) Bifurcation diagram showing primary circle pitchfork bifurcations from the conductive state to three branches containing five (TR$_5$, $Ra\geq Ra_5=9180$), six (TR$_6$, $Ra \geq Ra_6=9470$), and four (TR$_4$, $Ra\geq Ra_4=10124$) co-rotating rolls. The inset zooms in on the mixed-mode branch MM. Solid and dashed curves signify linearly stable and unstable states, respectively. (b) Schematic bifurcation diagram using a local quantity to distinguish symmetrically-related branches TR$_4$ and TR$_4^\prime\equiv\tau(L_x/8)$TR$_4$. MM connects branches TR$_4$ and TR$_4^\prime$ and also intersects branch TR$_6$ in a transcritical bifurcation at $Ra=10500$, whose eigenmodes $\pm e_{\rm trans}$ are shown in (c-d). (e-l) Evolution of temperature field along MM branch. Each roll contains a high-temperature red patch and a low-temperature blue patch. Selected rolls are shown as ovals with velocity indicated by arrows; all rotate in the same direction. Along pathway (e)-(f)-(g)-(h), two pairs of rolls (centred at $x\approx3$ and at $x\approx9$) merge, while along pathway (i)-(j)-(k)-(l), two single rolls (centred at $x\approx0$ and $x\approx6$) disappear. Traversed in the opposite direction from TR$_4$ and TR$_4^\prime$, these pathways correspond to roll splitting and roll creation. (m-n) Eigenmodes $e_{\rm split}$ and $e_{\rm create}$ responsible for pitchfork bifurcations from TR$_4$ and TR$_4^\prime$ at $Ra=10630$ and $Ra=10527$.}
\end{figure}

Figure \ref{BD_connector} shows the bifurcation structure obtained by numerical continuations of the Oberbeck--Boussinesq equations. We find three types of transverse roll states: TR$_5$, TR$_6$, and TR$_4$ containing five, six, and four identical co-rotating rolls. The bifurcation diagram of figure \ref{BD_connector}(a) shows that TR$_5$, TR$_6$, and TR$_4$ bifurcate sequentially from the conductive state at Rayleigh numbers of 9180, 9470, and 10124, respectively. Since TR$_4$ and TR$_6$ bifurcate after TR$_5$, they are unstable at onset, and also at all $Ra$ that we have investigated, as indicated by the dashed curves). The bifurcations are circle pitchforks, which produce a ``circle'' of states related by translation in the periodic $x$ direction. Figures \ref{BD_connector}(h) and (l) depict two TR$_4$ states that have different phases. The global quantity $\|\theta\|_2$ used to plot the bifurcation diagram of figure \ref{BD_connector}(a) does not distinguish between them. In order to do so, a schematic bifurcation diagram using an arbitrary local quantity is shown in figure \ref{BD_connector}(b).

A mixed-mode branch MM also appears in figure \ref{BD_connector}. The MM branch is produced by pitchfork bifurcations at $Ra=10527$ and $Ra=10630$ from the TR$_4$ states depicted in figures \ref{BD_connector}(h) and \ref{BD_connector}(l). Figure \ref{BD_connector}(b) shows that the mixed-mode branch connects TR$_4$ and TR$_4^{\prime}\equiv\tau(L_x/8)$TR$_4$. Figures \ref{BD_connector}(a) and (b) also show that MM intersects TR$_6$, a branch of six-roll states as depicted in figure \ref{BD_connector}(e) and (i). This intersection is a transcritical bifurcation and this is where we shall begin our detailed description of the mixed-mode branch.

\subsection{Tour of the mixed-mode branch: roll-merging and roll-disappearing}
We start our tour of the mixed-mode branch from figure \ref{BD_connector}(e) at $Ra=10500$, at which the connector branches have six rolls of equal size. This is the location of a transcritical bifurcation where the connector branches meet TR$_6$. Figures \ref{BD_connector}(c)-(d) show the eigenvector associated with this bifurcation. We first follow the path (e)-(f)-(g)-(h). At figure \ref{BD_connector}(f), a saddle-node bifurcation occurs, at which the inner halves of two roll pairs, at $x\approx3$ and $x\approx9$, weaken, allowing the outer halves to merge into wide diffuse rolls in figure \ref{BD_connector}(g). By figure \ref{BD_connector}(h), these have become identical to the remaining rolls, leading to a state with four rolls of equal size belonging to TR$_4$. The connector branch meets TR$_4$ in a pitchfork bifurcation at $Ra=10630$. We call this portion of the connector branch the {\it merging} pathway.

We return to figure \ref{BD_connector}(i) which shows the same temperature field as figure \ref{BD_connector}(e) and now traverse the path (i)-(j)-(k)-(l). Figure \ref{BD_connector}(j) shows the temperature field at a saddle-node bifurcation, at which the rolls centred at $x=6$ and $x=0$ weaken. At figure \ref{BD_connector}(k) they have further weakened and by figure \ref{BD_connector}(l), they have disappeared, again leading to a state with four rolls of equal size belonging to TR$_4^\prime$, a half-roll shifted version of TR$_4$. This intersection between the connector branch and  TR$_4^\prime$ at $Ra=10527$ is also a pitchfork bifurcation, but a different one. We call this portion of the connector branch the {\it disappearing} pathway. We emphasize that each of TR$_4$ and TR$_4^\prime$ contain both pitchfork bifurcations, at $Ra=10630$ and at $Ra=10527$. This is dictated by the fact that TR$_4$ and TR$_4^\prime$ are related by symmetry and is clearly shown in figure \ref{BD_connector}(b). 

We may equivalently traverse the connector branches in the opposite direction. Along the pathway (h)-(g)-(f)-(e), the rolls at $x=3$ and $x=9$ each {\it split}, while along (l)-(k)-(j)-(i), new rolls are {\it created} at $x=6$ and $x=0$. This allows us to interpret the eigenmodes $e_{\rm split}$ and $e_{\rm create}$ of TR$_4$ that we have shown in figures \ref{BD_connector}(m) and (n). Eigenmode $e_{\rm split}$, shown in figure \ref{BD_connector}(m), is responsible for the pitchfork bifurcation to the merging/splitting portion of the connector branch. When added to TR$_4$, its strong blue and red portions at $x=1.5$ and $4.5$ (and $x=7.5$, $10.5$) strengthen and move apart the outer parts of the roll centred at $x=3$ in figure \ref{BD_connector}(h). Splitting the blue and red portions of this roll leads to the formation of new complementary half-rolls in figure \ref{BD_connector}(f), to which the weaker portions of $e_{\rm split}$ surrounding $x=3$ also contribute. Eigenmode $e_{\rm create}$ of TR$_4^\prime$, shown in figure \ref{BD_connector}(n), is responsible for the pitchfork bifurcation to the disappearing/creating portion of the connector branch. When added to TR$_4^\prime$, its strong blue and red portions surrounding $x=6$ (and resembling a new roll) counteract the inner portions of the roll pair in the red rectangular box in figure \ref{BD_connector}(l). This new roll eventually parts the roll pair and inserts itself into the space in between, as shown in figure \ref{BD_connector}(k), (j). This also occurs simultaneously at $x=0$.

\subsection{Mode interactions and equivariant bifurcations}
The scenario reported here resembles that studied in \S 3.2.4--\S 3.2.5 of \citet{Zheng2024part1}. The crucial feature distinguishing the two cases is that here, all of the states in figures \ref{BD_connector}(c-n) have two sub-intervals $[0,L_x/2]$ and $[L_x/2,L_x]$ in which the states are identical. This means that the entire bifurcation scenario can be considered as taking place in $[0,L_x/2]$, where the number of rolls is not four and six, but two and three. This leads to the scenario of $D_2$--$D_3$ mode interaction, whereas \citet{Zheng2024part1} concerned the interaction of branches with $D_3$ and $D_4$ symmetry.

The mathematical framework and equations 
which we will adapt to our configuration are presented in \S 3.3 of \citet{dangelmayr1986steady}. 
We consider a system with $O(2)$ symmetry in which modes with wavenumbers $m$ and $n$ bifurcate simultaneously or close together. Because the conductive state is uniform in the periodic direction $x$, its eigenvectors are trigonometric functions of $x$ and the eigenvectors are of the form ${\bf f}_m(z) e^{imx (4\pi/L_x)}$ and ${\bf f}_n(z)e^{inx(4\pi/L_x)}$, where ${\bf f}_m$ and ${\bf f}_n$ contain the temperature and velocity components and depend on the remaining cross-channel spatial component $z$. The factor $4\pi/L_x=2\pi/(L_x/2)$ comes from the fact that the period in $x$ for our case is $L_x/2$ rather than $2\pi$. 

Solutions (deviations from the conductive state) in the vicinity of the bifurcations are approximated as a sum of the two eigenvectors: 
\begin{align}
	z_m(t) e^{im x(4\pi/L_x)} {\bf f}_m + z_n(t) e^{inx(4\pi/L_x)} {\bf f}_n
	+ \mbox{c.c.},
	\label{eq:definez}
\end{align}
weighted by complex time-dependent amplitudes $z_m$ and $z_n$. (The wall-normal spatial coordinate $z$ will no longer appear and the reflection operator $\pi_{xz}$ will be considered here only as a reflection in $x$.) The action of $O(2)$ on \eqref{eq:definez} is given via its generators: 
\vspace*{-0.1cm}
\begin{subequations}
	\label{eq:symspace}
	\begin{align}
		\mbox{Reflection in}~x:~& z_m(t) e^{-im x(4\pi/L_x)} {\bf f}_m + z_n(t) e^{-inx(4\pi/L_x)} {\bf f}_n + \mbox{c.c.}, \label{eq:symspace_a}\\
		\mbox{Translation by}~x_0:~& 
		z_m(t) e^{im (x+x_0)(4\pi/L_x)} {\bf f}_m + z_n(t) e^{in(x+x_0)(4\pi/L_x)} {\bf f}_n+ \mbox{c.c.}. \label{eq:symspace_b}
	\end{align}
\end{subequations}
This action can therefore be transferred to $z_m$ and $z_n$ as:
\vspace*{-0.1cm}
\begin{subequations}
	\label{eq:symgen}
	\begin{align}
		\mbox{Reflection in}~x:~& (z_m,z_n) \rightarrow (\bar{z}_m,\bar{z}_n), \label{eq:symgen_a}\\
		\mbox{Translation by}~x_0:& ~(z_m,z_n) \rightarrow (z_m e^{imx_0(4\pi/L_x)}, z_n e^{i
			nx_0(4\pi/L_x)}). \label{eq:symgen_b}
	\end{align}
\end{subequations}
In the same way that symmetry operations can be transferred from fields \eqref{eq:symspace} to amplitudes \eqref{eq:symgen}, the time evolution of a field is to be represented by a system of evolution equations for $z_m$ and $z_n$. This system of equations must be {\it equivariant}, i.e.\ it must commute with all of the operators in the symmetry group. The most general such evolution equations are given by \citet{dangelmayr1986steady} and \citet{dangelmayr_Armbruster_1986} as: 
\vspace*{-0.1cm}
\begin{subequations}
	\label{eq:normalz}
	\begin{align}
		\dot{z}_m&=p_m z_m + q_m\bar{z}_m^{n-1} z_n^m, \\
		\dot{z}_n&=p_n z_n + q_n\bar{z}_m^n z_n^{m-1},
	\end{align}
\end{subequations}
where $p_m, p_n, q_m, q_n$ are polynomial functions of the {\it invariants} $|z_m|^2$, $|z_n|^2$, and $z_m^n\bar{z}_n^m+\bar{z}_m^nz_n^m$ and of two parameters: a bifurcation parameter $\mu$ and an unfolding parameter $\alpha$ measuring the distance between the threshold of modes $z_m$ and $z_n$. Introducing polar coordinates $z_m=r_me^{i\theta_m}$, $z_n=r_ne^{i\theta_n}$, $\theta\equiv n\theta_m-m\theta_n$, the evolution equations \eqref{eq:normalz} are rewritten as:
\vspace*{-0.1cm}
\begin{subequations}
	\begin{align}
		\dot{r}_m&=p_mr_m + q_m r_m^{n-1}r_n^m \cos\theta,\\
		\dot{r}_n&=p_nr_n + q_n r_m^n r_n^{m-1} \cos\theta,\\
		\dot{\theta}&=-(nq_mr_n^2+mq_n r_m^2) \: r_m^{n-2} r_n^{m-2}\sin\theta.
	\end{align}
\end{subequations}
The general functions $p_m, p_n, q_m, q_n$ are then Taylor expanded and truncated, and rescalings are performed to normalize the largest possible number of coordinates, leading to:
\vspace*{-0.1cm}
\begin{subequations}
	\begin{align}
		p_m&=-\nu(\epsilon_m r_m^2 + \rho r_n^2 -\mu + \alpha), 
		&q_m&=-\nu\beta_m, \\ 
		p_n&=-(\epsilon_n r_n^2 + \kappa r_m^2 -\mu),&
		q_n&=\beta_n,  
	\end{align}
\end{subequations}
where $\epsilon_{m,n}=\pm 1$, at least one of $\beta_m$, $\beta_n$ is non-zero, $\nu>0$, and various other conditions are required of the other coefficients. (This form excludes the interesting special case $(m,n)=(1,2)$, for which $\epsilon_m=\kappa=0$.)

We now specialize to our case, for which $(m,n)=(2,3)$. The governing equations become:
\vspace*{-0.5cm}
\begin{subequations}
	\label{eq:Smn}
	\begin{align}
		\dot{r}_2/\nu & = -(\epsilon_2 r_2^2 + \rho r_3^2 -\mu + \alpha)r_2  -\beta_2 r_2^2 r_3^2 \cos\theta\\
		\dot{r}_3 & =-(\epsilon_3 r_3^2+\kappa r_2^2 -\mu)r_3 +\beta_3 r_2^3 r_3 \cos\theta \\
		\dot{\theta}& =-(-3\nu \beta_2r_3^2+2\beta_3 r_2^2) \: r_2 \sin\theta.
	\end{align}
\end{subequations}
Obvious steady solution to \eqref{eq:Smn} are the trivial solution $r_2=r_3=0$ and the two pure-mode solutions $S_2$ and $S_3$:
\vspace*{-0.2cm}
\begin{subequations}
	\label{eq:S23}
	\begin{align}
		&S_2: r_2= \sqrt{(\mu - \alpha)/\epsilon_2} &&\qquad r_3=0 \qquad \mbox{with bifurcation point} \quad \mu=\mu_2=\alpha, \label{eq:bifS2}\\
		&S_3: r_3=\sqrt{\mu/\epsilon_3} &&\qquad r_2=0 \qquad \mbox{with bifurcation point} \quad \mu=\mu_3=0. \label{eq:bifS3}
	\end{align}
\end{subequations}
The bifurcations creating these are circle pitchforks, in which $S_2$ and $S_3$ can have any phases $\theta_2, \theta_3\in [0,2\pi)$. Applying a phase shift is equivalent to a translation in $x$; see \eqref{eq:symspace_b} and \eqref{eq:symgen_b}. A half-wavelength shift, as in TR$_4 \rightarrow$ TR$_4^\prime$ or TR$_6 \rightarrow$ TR$_6^\prime$, is equivalent to shifting $\theta_{2,3}$ by $\pi$ or, equivalently, multiplying $r_{2,3}$ by $e^{i\pi}=-1$. There exist two more steady solutions $S_+$ and $S_-$ for which $r_2\neq 0$, $r_3\neq 0$, satisfying: 
\vspace*{-0.15cm}
\begin{subequations}
	\label{eq:Spm}
	\begin{align}
		&\sin\theta = 0 \Longrightarrow \cos\theta=\pm 1,\nonumber\\
		S_+:& \left\{\begin{array}{l} 0 = -(r_2^2 + \rho r_3^2 -\mu + \alpha) -\beta_2 r_2r_3^2 \\
			0 =-(\kappa r_2^2 + r_3^2 -\mu) +\beta_3 r_2^3 \end{array}\right\},\label{eq:Sp}\\
		S_-:& \left\{\begin{array}{l} 0 =  -(r_2^2 + \rho r_3^2 -\mu + \alpha) +\beta_2 r_2r_3^2 \\
			0 =-(\kappa r_2^2 + r_3^2 -\mu)  -\beta_3 r_2^3 \end{array}\right\}.\label{eq:Sm}
	\end{align}
\end{subequations}
There also exist travelling wave solutions, satisfying $\sin\theta\neq 0$, which we will not discuss.

\subsection{Fitting the parameters}
\begin{figure}
	\centering
	\captionsetup{font={footnotesize}}
	\captionsetup{width=13.5cm}
	\captionsetup{format=plain, justification=justified}
	\includegraphics[width=\columnwidth]{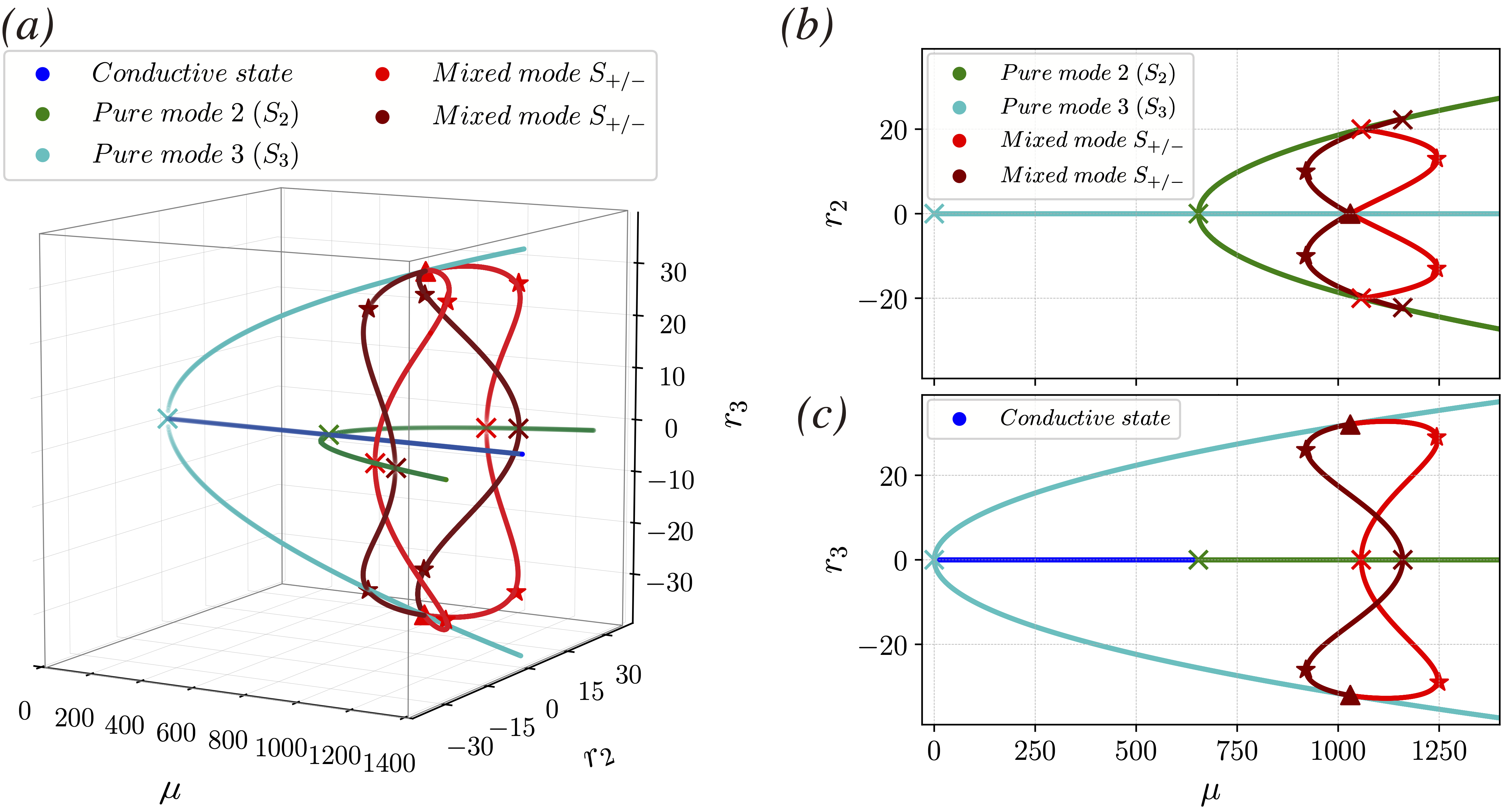}
	\caption{\label{D2D3_DA} (a) Solution curves in the $D_2$--$D_3$ mode interaction scenario, and in two-dimensional projections (b-c). The trivial solution $r_2=r_3=0$ corresponds to the hydrodynamic conductive state, the pure-mode solutions \eqref{eq:S23} correspond to the four-roll branch TR$_4$ (green) and the six-roll branch TR$_6$ (cyan), and the mixed-mode solutions \eqref{eq:Spm} correspond to the merging (brown) and disappearing (red) portions of the mixed-mode (connector) branch. Pitchfork, saddle-node, and transcritical bifurcation points are marked by crosses, stars, and triangles, respectively. Intersections not marked by a symbol are merely projection effects.}
\end{figure}

The bifurcations that we have observed numerically in our hydrodynamic simulations will allow us to determine the values of all of the parameters for our case. We will then be able to confirm the correspondence of this general symmetry-dictated scenario with our observations, culminating in figure \ref{D2D3_DA}, which we will describe at the end of this subsection. The versions of $S_2$ and $S_3$ that we observe in our hydrodynamic system are the restrictions to the interval $[0,L_x/2]$ of the branches that we have called TR$_4$ and TR$_6$ in figure \ref{BD_connector}. The bifurcations creating them are both supercritical, so $\epsilon_2=\epsilon_3=1$. The bifurcation to $S_3$ occurs at $Ra=9470$ so, according to \eqref{eq:bifS3}, $\mu = Ra-9470$. The bifurcation to $S_2$ occurs at $Ra=10\,124$, implying via \eqref{eq:bifS2} that $\alpha= 10\,124-9470=654$. 

We calculate the points at which branches $S_\pm$ bifurcate from $S_3$. Since $r_2=0$, the last terms in \eqref{eq:Sp} and \eqref{eq:Sm}, which differ between the two, vanish and so $S_+$ and $S_-$ bifurcate simultaneously from $S_3$. The condition for this is:
\vspace*{-0.1cm}
\begin{align} 
	S_3 ~\mbox{meets}~ S_{\pm}:& ~~
	0 = -( \rho \mu -\mu + \alpha)\Longrightarrow \mu=\alpha/(1-\rho).
\end{align}
In our hydrodynamic calculations, this bifurcation occurs at $Ra=10500$; see figure \ref{BD_connector}(a,b,e,i). This Rayleigh number corresponds to $\mu=1030$, implying that $(1-\rho)=\alpha/\mu= 654/1030=0.635$ and hence that $\rho=0.365$. The bifurcation is transcritical, with $S_+$ and $S_-$ branching from $S_3$ in opposite directions, as seen in figure \ref{BD_connector}(a-b). This is typical for bifurcations breaking $D_3$ symmetry; see \cite{Zheng2024part1}.

We now calculate the points at which $S_\pm$ bifurcate from $S_2$:
\vspace*{-0.2cm}
\begin{subequations}
	\label{eq:S2meetsSpm}
	\begin{align}
		S_2 ~\mbox{meets}~ S_+:& ~~
		0 =-(\kappa (\mu-\alpha) -\mu) +\beta_3 (\mu-\alpha)^{3/2} ,\\
		S_2 ~\mbox{meets}~ S_-:& ~~
		0 =-(\kappa (\mu-\alpha) -\mu) -\beta_3 (\mu-\alpha)^{3/2}. 
	\end{align}
\end{subequations}
Each of these meeting points is a pitchfork bifurcation. In our case, these points are at $Ra=10630$ and $Ra=10527$, as shown in figure \ref{BD_connector}(a,b,h,l). This leads to $\mu\sim 1160$, $\mu-\alpha=506$ and $\mu=1057$, $\mu-\alpha=403$. Substituting these values into \eqref{eq:S2meetsSpm}, we obtain $\kappa = 2.467$ and $\beta_3=-0.00776$. 
The remaining unknown parameter in \eqref{eq:Spm} is $\beta_2$, which we determined by fitting the location of one of the two saddle-node bifurcations in our hydrodynamic system. One saddle-node bifurcation occurs at $Ra=10714$, corresponding to $\mu=1244$. By manually tuning $\beta_2$ in \eqref{eq:Spm} to produce a saddle-node bifurcation at this location, we obtained $\beta_2=0.012$. 

With the set of parameter values now complete, we sought a second saddle-node bifurcation in \eqref{eq:Spm} which would match the remaining saddle-node bifurcation in our hydrodynamic system at $Ra=10356$, corresponding to $\mu=886$. Indeed, we located a second saddle-node bifurcation at $\mu=915$, fairly close to 886. The locations of the other bifurcations (pitchfork, transcritical, and first saddle-node) from the normal form and from our simulation match by construction; agreement was assumed and used to determine the normal-form parameters. However, the existence of the second saddle-node bifurcation in the normal form and the fairly close agreement with its hydrodynamic value constitute evidence that the normal-form equations correctly describe the hydrodynamic simulation in this Rayleigh-number range.

We now assemble all of these results into figure \ref{D2D3_DA}, which depicts the solution branches $S_2$, $S_3$, and $S_\pm$ that we have computed from \eqref{eq:S23} and \eqref{eq:Spm} using the parameters that we have deduced from our hydrodynamic results. Each parabola represents two phases of solutions $S_2$ (green) and $S_3$ (cyan), with each half-parabola corresponding to a half-wavelength translation of the other half. Note that we have chosen to allow negative values for $r_2$ and $r_3$, which avoids non-smoothness of the curves at the $r_2=0$ and $r_3=0$ axes. With this convention, $\theta_2$ and $\theta_3$ are restricted to $[0,\pi)$ rather than to $[0,2\pi)$.

Branches $S_\pm$ together comprise our mixed-mode connector branch. We have coloured these by whether they correspond to the ``merging'' (brown, figure \ref{BD_connector}(e,f,g,h)) or ``disappearing'' (red, figure \ref{BD_connector}(i,j,k,l)) half-branch; see figure \ref{BD_connector}(a,b), where we have used the same colour-coding for the connector branch. Each of the merging and disappearing branches begins at one half-parabola of $S_3$ (in a transcritical bifurcation), undergoes a saddle-node bifurcation, then intersects a half-parabola of $S_2$ (in a pitchfork bifurcation), and ends on the other half-parabola of $S_3$. It is tempting to associate one of $S_+$, $S_-$ with the merging half-branch and the other with disappearing, but this is the case only if $r_2$, $r_3$ are required to be positive, the convention opposite to that we have chosen for figure \ref{D2D3_DA}.

\subsection{Comparison with $D_3$--$D_4$ scenario}
\cite{Zheng2024part1} discussed bifurcation from a branch which is very similar to the TR$_4$ branch investigated currently: with four identical rolls, these states have $D_4$ symmetry. This case has many features in common with the current one, namely a connector branch between pure-mode branches. The connector branch exists between a transcritical bifurcation with a three-roll branch and pitchfork bifurcations to four-roll branches of different phases. The connector branch has two pathways, one in which rolls merge or split and the other in which rolls disappear or are created.

The present case is slightly different. Here, the eigenvectors which bifurcate from TR$_4$ to create the connector branch, shown in figures \ref{BD_connector}(m) and (n), have the additional half-domain translation symmetry $\tau(L_x/2)$. 
Consequently, the connector branch created by these bifurcations also has this symmetry. 
The relevant symmetry becomes $D_2$ rather than $D_4$.
The pitchfork bifurcations associated with the two portions of the connector branch (merging/splitting and disappearing/creating) occur at different Rayleigh numbers, unlike the single pitchfork bifurcation in the $D_4$ case which produced both portions of the connector branch.
An analogy can be drawn to the study by \cite{bergeon2001three} of Marangoni-B\'enard convection. In a square container, a single bifurcation gives rise to two sets of non-symmetrically related (and therefore non-equivalent) convective branches \citep{swift1985bifurcation}, taking the form of straight or diagonal states. 
As the domain was modified from square to slightly rectangular, the simultaneous bifurcations of the $D_4$ case split into two. 

The other pure-mode branch in \cite{Zheng2024part1} was a three-roll state, with $D_3$ symmetry, whose normal form leads to a three-fold transcritical bifurcation \citep{gambaudo1985perturbation, golubitsky1988singularities, Dawes2005}. We recover the same transcritical bifurcation here because the TR$_6$ state acts like two identical copies of a three-roll state.

\section{Conclusion}
\label{sec:conclusion}
The normal-form equations for $D_2$--$D_3$ interaction are well known, and so is the hydrodynamic configuration of inclined-layer convection. What we have done is to combine the two, showing that these general equations based only on symmetry considerations accurately describe numerical simulations of our specific problem. Computing the normal form coefficients from the  bifurcations observed in simulations of our hydrodynamic system, we have achieved close agreement between these two very different approaches.

More significantly, we have brought the bifurcations and branches of the normal form to life. The two mixed-mode branches distinguished quantitatively by a phase ($0$, $\pi$) or a sign ($\cos\theta =\pm 1$) in the equations turn out to be manifestations of two phenomenologically different routes for changing the numbers of rolls: first, the merging of two rolls into one (or, reversing the direction, the splitting of one roll into two) and second, the disappearance (or creation) of a single roll. Saddle-node bifurcations are indeed turning points in the sizes of the rolls, occurring at states in which a roll is too small to survive, or a pair of rolls too weak to prevent the formation of a roll between them. Eigenvectors focus in on the spatial location at which these transitions occur. This is the essence of the famous statement by \cite{poincare1908avenir}, the founder of the modern theory of dynamical systems: {\it ``Mathematics is the art of giving the same name to different things''}.

\backsection[Acknowledgements]{The authors wish to acknowledge Edgar Knobloch for inspiring this investigation.}
\backsection[Funding]{This work was supported by the European Research Council (ERC) under the European Union's Horizon 2020 research and innovation programme (grant no. 865677).}
\backsection[Declaration of interests]{The authors report no conflict of interest.}

\bibliographystyle{jfm}

\end{document}